\documentclass[twocolumn,showpacs,showkeys,prd]{revtex4}
\usepackage{graphicx}

\def\GeV{\hbox{ GeV}}

\begin{document}
%%%%%%%%%%%%%%%%%%%%%%%%%%%%%%%%%%%%%%%%%%%%%%%%%%%%%%%%%%%%%%%%%%%%%%

\title{Constraining an $R$-parity violating supergravity
  model with the Higgs induced \\Majorana neutrino magnetic moments}

\author{Marek G\'o\'zd\'z}
\email{mgozdz@kft.umcs.lublin.pl}

\affiliation{
Department of Informatics, Maria Curie-Sk{\l}odowska University, \\
pl.~Marii Curie--Sk{\l}odowskiej 5, 20-031 Lublin, Poland}

\begin{abstract}
  It is well known, that $R$-parity violating supersymmetric models
  predict a~non-zero magnetic moment for neutrinos. In this work we
  study the Majorana neutrino transition magnetic moments within an RpV
  modified minimal supergravity model. Specifically, we discuss the
  contributions coming from the charged Higgs bosons, higgsinos,
  leptons, sleptons, charged gauge bosons, and charginos. We use the
  experimental results from the MUNU collaboration to restrict the
  model's parameter space. A~comparison with two other types of
  contributions (only trilinear RpV couplings and trilinear plus
  neutrino--neutralino mixing) is also presented. We have found that the
  presently discussed processes dominate significantly, exceeding in
  some cases even the experimental limits.
\end{abstract}

\pacs{11.30.Pb, 12.60.Jv, 14.60.Pq} 
\keywords{neutrino magnetic moment, supersymmetry, R-parity}

\maketitle

%%%%%%%%%%%%%%%%%%%%%%%%%%%%%%%%%%%%%%%%%%%%%%%%%%%%%%%%%%%%%%%%%%%%%%%%
\section{Introduction}

Neutrinos are one of the most mysterious particles known. They interact
very weakly with matter, taking part only in the weak (and presumably
gravitational) interactions. Due to mixing \cite{fogli} their mass
eigenstates and interaction eigenstates differ substantially, which
gives rise to the generation lepton number violating oscillations
\cite{nu-osc}. They are also the lightest fermions ever observed -- in
fact the smallness of neutrino masses cannot be explained using the
standard Higgs mechanism \cite{nu-mass}. All this indicates, that
neutrinos are a window to some new physics beyond the standard model.

The possible electromagnetic properties of neutrinos in the form of an
induced magnetic moment are particularly interesting, since the
electromagnetic interactions are much easier to control in any precise
measurement. This problem has been studied in the standard model
\cite{kayser82,vogel99} and its extensions. In Ref.~\cite{magmom-lnv}
the magnetic moments have been reviewed in models with spontaneously
broken lepton number. A~lot of work has been devoted to discuss the
neutrino masses in the $R$-parity violating (RpV) supersymmetry
\cite{Haug,Bhatta,Abada,Grossman,mg06,mg,mg17,mg18}. In these models, the
neutrino mass is generated mostly at the one-loop level through an RpV
loop containing in the simplest version a~particle--sparticle pair
(lepton--slepton or quark--squark). The approach usually used in the
literature assumed no mixing among particles, and some average common
value for the masses of the superparticles at low energies. In
Ref.~\cite{dl} a~basis independent classification of various diagrams
leading to Majorana neutrino masses has been presented. The study of
neutrino masses can be directly extended to the magnetic moment case by
introducing an interaction with an external photon.

A~more systematic approach to the calculations of the magnetic moments
has been presented in our previous works \cite{mg09,mg19,mg17,mg20}. The
SUGRA model has been used to calculate masses and magnetic moments in
Ref.~\cite{mg09}. The contribution coming from neutrino--neutralino
mixing has been discussed in Ref.~\cite{mg19}, while the contribution
from generic diagrams with two mass insertions on the internal lines was
presented in \cite{mg17}. Calculations within the gauge-mediated model
were performed in \cite{mg20}. We have studied the Majorana neutrino
masses and transition magnetic moments using exact analytic formulas and
up-to-date experimental data concerning neutrino oscillations and the
non-observability of the neutrinoless double beta decay
($0\nu2\beta$). Our numerical procedure rested upon the assumption that
most of the free parameters of the supersymmetric model unify in certain
way at the GUT scale $m_{\rm GUT}\sim 1.2\times 10^{16}\GeV$. This
allows to reduce the number of free parameters to few, and calculate the
rest by performing renormalization group evolution. Such approach gives
a~much more convincing picture of the low energy spectrum than just
assessing the running masses of the superparticles. We have also
demonstrated the influence of the mixing among squarks and quarks on the
results. The evident weak point of our approach was the lack of
knowledge of the trilinear coupling constants, which forced us to use
additional source of constraints (neutrino oscillations, $0\nu2\beta$
decay).

In this paper we study the Majorana neutrino transition magnetic moments
generated in the $R$-parity violating modified minimal supergravity
model. This model, based on the Ref.~\cite{mSUGRA}, takes into account
the full dependence of the renormalization group equations (RGE) on the
RpV couplings and their soft SUSY breaking versions. All these couplings
have been incorporated in the unification scheme, so, within this
assumption, the whole model can be calculated numerically. No other
input, in particular experimental data, is needed. We concentrate on the
contributions coming from the charged Higgs bosons, charged higgsinos,
charged gauge bosons, charginos, leptons, and sleptons. The neutral
particles, like the neutral components of the Higgs bosons, neutralinos
and sneutrinos will contribute to the neutrino mass, but will not
contribute to the magnetic moment, and as such will not be discussed
here.

The paper is organized as follows. After the introduction, in Sect.~II
the minimal supersymmetric standard model with $R$-parity violation and
supersymmetry broken by the gravitational interactions is
presented. Sect.~III is devoted to the technical problems of handling
the free parameters and calculating the coupling constants in the Higgs
sector during the electroweak symmetry breaking. The Majorana neutrino
transition magnetic moments, and the interesting for us contributions to
this quantity are discussed in Sect.~IV. At the end numerical results
are presented.

\section{$R$-parity violating SUGRA}

The minimal supersymmetric standard model (MSSM) \cite{mssm,kazakov} is
the minimal extension of the standard model (SM) of particles and
interactions, which introduces supersymmetry (SUSY). In short, for each
particle there is a~new superpartner (fermionic for bosons and bosonic
for fermions) so that the number of fermionic and bosonic degrees of
freedom in the MSSM are equal. The one additional multiplet, not present
in the SM, is a~second Higgs doublet needed to cancel the gauge
anomalies and to give masses to both the up and down components of the
$SU(2)$ doublets. This version of the model preserves the so-called
$R$-parity, which is a~multiplicative quantum number defined as
$R=(-1)^{2S+3(B-L)}$, $S$ being the spin of the particle, $B$ the
baryon, and $L$ the lepton number. One easily sees, that $R=+1$ for
ordinary standard model particles, while $R=-1$ for the supersymmetric
partners. Preserving $R$ not only gives the baryon and lepton number
conservations but also forbids decays of SUSY particles into SM
particles, leaving the lightest SUSY particle stable.

The problem of the lepton and baryon number conservation is still
disputable. On one hand, our observations suggest, that both $B$ and $L$
are conserved quantum numbers. Especially strong limits on $B$ violation
come from the non-observation of the proton decay. On the other hand, in
the SM not only the total lepton number but also generation lepton
numbers $L_e$, $L_\mu$, and $L_\tau$ are conserved, and this rule has
been proven wrong after the discovery of neutrino oscillations. From the
theoretical point of view, there is no underlying symmetry or mechanism,
which supports conservation of $B$ and $L$, which might suggest that it
is only an accidental (not exact) symmetry present in the low-energy
regime. It is generally expected, that the higher-energy extensions of
the SM will not preserve at least the lepton number (the proton
stability must still hold). Such models based on supersymmetry introduce
new interactions, or in fact do not rule out certain terms, that should
be present in the superpotential, which violate $R$-parity
\cite{barbier,aul83,valle,np,rbreaking}. They have a~much richer
phenomenology and predict lots of new phenomena, like the lepton number
violating neutrinoless double beta decay, RpV loop-corrected neutrino
mass and magnetic moment, and others.

We work within the slightly modified model described in details in
Ref.~\cite{mSUGRA}. It is defined by the following $R$-parity conserving
and $R$-parity violating parts of the superpotential
\begin{equation}
  W = W_{\rm RpC} + W_{\rm RpV},
\label{W}
\end{equation}
where
\begin{eqnarray}
  W_{\rm RpC}&=& \epsilon_{ab} \Big[ ({\bf Y}_E)_{ij} L_i^a
  H_d^b {\bar E}_j + ({\bf Y}_D)_{ij} Q_i^{ax} H_d^b {\bar D}_{jx} 
  \nonumber \\
  &+&
  ({\bf Y}_U)_{ij} Q_i^{ax} H_u^b {\bar U}_{jx} 
  - \mu H_d^a H_u^b \Big], 
  \label{WRpC}\\
  W_{\rm RpV}&=&
  \epsilon_{ab}\left[ 
    \frac{1}{2}({\bf \Lambda}_{E^k})_{ij} L_i^a L_j^b{\bar E}_k +
    ({\bf \Lambda}_{D^k})_{ij} L_i^a Q_j^{xb} {\bar D}_{kx} \right]
  \nonumber \\
  &+&
  \frac{1}{2}\epsilon_{xyz} ({\bf \Lambda}_{U^i})_{jk} {\bar U}_i^x{\bar
    D}_j^y{\bar D}^z_k - \epsilon_{ab} \kappa^i L_i^a H_u^b.
  \label{WRpV}
\end{eqnarray}
Here {\bf Y}'s are the 3$\times$3 trilinear Yukawa-like couplings, $\mu$
the bilinear Higgs coupling, and $\bf \Lambda$ and $\kappa^i$ represent
the magnitudes of the $R$-parity violating trilinear and bilinear
terms. $L$ and $Q$ denote the $SU(2)$ left-handed doublets, while $\bar
E$, $\bar U$ and $\bar D$ are the right-handed lepton, up-quark and
down-quark $SU(2)$ singlets, respectively. $H_d$ and $H_u$ mean two
Higgs doublets. We have introduced color indices $x,y,z = 1,2,3$,
generation indices $i,j,k=1,2,3=e,\mu,\tau$ and the $SU(2)$ gauge
indices $a,b = 1,2$. Finally, $\epsilon_{ab}$ and $\epsilon_{xyz}$ with
$\epsilon_{12}=\epsilon_{123}=1$ are the totally antisymmetric tensor
densities.

The supersymmetry is not observed in the regime of energies accessible
to our experiments. Therefore it must be broken at some
point. A~convenient method to take this fact into account is to
introduce explicit terms, which break supersymmetry in a~soft way,
\textit{ie.}, they do not suffer from ultraviolet divergencies. We add
them in the form of a~scalar Lagrangian \cite{mSUGRA},
%
%%%%%%%%%%%%%%%%%%%%
\begin{widetext}
%%%%%%%%%%%%%%%%%%%%
\begin{eqnarray}
  -{\cal L} &=&
  m_{H_d}^2 h_d^\dagger h_d + m_{H_u}^2 h_u^\dagger h_u
  + l^\dagger ({\bf m}_L^2) l 
  + {l_i}^\dagger ({\bf m}_{L_i H_d}^2) h_d
  + h_d^\dagger ({\bf m}_{H_d L_i}^2) l_i 
  + q^\dagger ({\bf m}_Q^2) q + e ({\bf m}_E^2) e^\dagger
  +  d ({\bf m}_D^2) d^\dagger + u ({\bf m}_U^2) u^\dagger \nonumber \\
  &+& \frac12 \left( 
    M_1 \tilde{B}^\dagger \tilde{B} + 
    M_2 \tilde{W_i}^\dagger \tilde{W^i} +
    M_3 \tilde{g_\alpha}^\dagger \tilde{g^\alpha} + {\rm h.c.}\right )
  \\
  &+& [ ({\bf A}_E)_{ij} l_i h_d e_j
  +    ({\bf A}_D)_{ij} q_i h_d d_j
  +    ({\bf A}_U)_{ij} q_i h_u u_j 
  -   B h_d h_u                \nonumber \\ 
  &+&  ({\bf A}_{E^k})_{ij} l_i l_j e_k
  +    ({\bf A}_{D^k})_{ij} l_i q_j d_k
  +    ({\bf A}_{U^i})_{jk} u_i d_j d_k               
  -   D_i l_i h_u + {\rm h.c.} ], \nonumber
\end{eqnarray}
%%%%%%%%%%%%%%%%%%%%
\end{widetext}
%%%%%%%%%%%%%%%%%%%%
% 
where the lower case letter denotes the scalar part of the respective
superfield. $M_i$ are the gaugino masses, and $\bf A$ ($B$, $D_i$) are
the soft supersymmetry breaking equivalents of the trilinear (bilinear)
couplings from the superpotential.

\section{Free parameters and the low-energy spectrum}

The main problem of most supersymmetric models is their large number of
free parameters. Therefore additional constraints are usually imposed,
which introduce certain relations among them, effectively reducing their
number. It is a~well known fact, that the RGE equations for the gauge
coupling constants lead in the MSSM model to their unification at energy
$m_{\rm GUT}\approx 1.2\times 10^{16}\GeV$. This feature is absent in
the SM, although suggested by the extrapolation of the LEP1 data. It may
therefore seem natural to assume also certain type of unification of
other parameters. Let us introduce at $m_{\rm GUT}$:
\begin{itemize}
\item the common mass of all the scalars $m_0$,
\item the common mass of all the fermions $m_{1/2}$,
\item the common proportionality factor $A_0$ for the soft SUSY breaking
  couplings
  \begin{eqnarray}
    && {\bf A}_{U,D,E} = A_0 {\bf Y}_{U,D,E}, \nonumber\\
    && {\bf A}_{U^i,D^i,E^i} = A_0 {\bf Y}_{U,D,E} \quad (i=1,2,3).
  \end{eqnarray}
\end{itemize}
This scheme of unification follows the mSUGRA assumptions with the
exception that we do not assume the total universality of the ${\bf A}$
couplings, but vary them keeping them proportional to the Yukawa ${\bf
  Y}$ coupling constants at the unification scale. As a~free parameter
remains also the ratio of the Higgs vacuum expectation values,
$\tan\beta=v_u/v_d$, and the sign of the $\mu$ coupling constant, ${\rm
  sgn}(\mu)$. The sixth free parameter in our considerations is the
initial value of the $\bf \Lambda$'s at $m_Z$, $\Lambda_0$. We leave
this value free and investigate its influence on the results. The idea
is that we want to have $\bf \Lambda$'s non-zero and contributing from
the beginning to the RGE running of other parameters. In this way,
$\Lambda_0$ controls the amount of $R$--parity violation in the low
energy regime. Notice, that $\Lambda$'s will not unify at $m_{\rm GUT}$,
although their RGE running is almost flat (cf. Fig.~1 in
Ref.~\cite{mg01}).

The procedure of obtaining the low energy spectrum of the model involves
few renormalization group runnings between the low ($m_Z$) and high
energy scale ($m_{\rm GUT}$), and between $m_Z$ and the scale at which
the electroweak symmetry breaking occurs, which is defined by the top
squark mass eigenstates as
\begin{equation}
  q_{\rm min}=\sqrt{m_{\tilde t_1} m_{\tilde t_2}}.
\label{qmin}
\end{equation}
At this scale the radiative corrections to the Higgs potential coming
from the squarks are minimal. We start by fixing the Yukawa couplings at
$m_Z$ using the quark and lepton mass matrices $M_{U,D,E}$
\begin{eqnarray}
M_U &=& v_u \mathbf{S}_{U_R} \mathbf{Y}_U^T
\mathbf{S}_{U_L}^\dagger, \nonumber \\
M_D &=& v_d \mathbf{S}_{D_R} \mathbf{Y}_D^T
\mathbf{S}_{D_L}^\dagger, \\
M_E &=& v_d \mathbf{S}_{E_R} \mathbf{Y}_E^T
\mathbf{S}_{E_L}^\dagger, \nonumber
\end{eqnarray}
where {\bf S} matrices perform diagonalization so that one obtains
eigenstates in the mass representation. The Yukawa-like RpV couplings at
$m_Z$ are all set to
\begin{equation}
\mathbf{\Lambda} = \Lambda_0 {\bf 1},
\end{equation}
$\bf 1$ being a~$3\times 3$ unit matrix. Next, we evolve the gauge and
Yukawa couplings from $m_Z$ to the unification scale.  During this
running we turn off the Higgs sector, $\mu=\kappa_i=B=D_i=0$, and set
the sneutrino vev's $v_i=0$. These parameters will be calculated later.

At $m_{\rm GUT}$ we impose the GUT conditions as described above, ie.,
we set the massess of all the scalars to $m_0$, of all the fermions to
$m_{1/2}$, and the soft breaking couplings to $A_0 {\bf Y}$. The Yukawa
couplings ${\bf Y}$'s are left unchanged in order to reproduce correctly
the masses of leptons and quarks when evolved down to $m_Z$. So are the
${\bf \Lambda}$'s, which are free parameters here. We evolve all the
running parameters, coupling constants and masses, down, and find the
best scale for minimizing the scalar potential Eq.~(\ref{qmin}) and at
this scale calculate $\mu={\rm sgn}(\mu) \sqrt{|\mu|^2}$ and $B$ using
\cite{mSUGRA}
\begin{eqnarray}
  |\mu|^2 &=& 
  \frac{m_{H_d}^2 - m_{H_u}^2 \tan^2\beta}{\tan^2\beta -1} -
  \frac{M_Z^2}{2}, \\
  B &=& \frac{\sin 2\beta}{2} (m_{H_d}^2 - m_{H_u}^2 + 2|\mu|^2).
\end{eqnarray}
Running all the parameters down $q_{\rm min} \to m_Z$ non-zero $\mu$ and
$B$ generate non-zero $\kappa_i$ and $D_i$. These are used as the
starting values to the second long run. Next, we perform once again the
RGE running up to the $m_{\rm GUT}$ scale, but now with all the Higgs
parameters contributing to the running. We impose GUT conditions, go
down, calculate the new $q_{\rm min}$ scale and once again evaluate the
corrected Higgs parameters, this time together with the sneutrino vev's,
according to

%%%%%%%%%%%%%%%%%%%%
\begin{widetext}
%%%%%%%%%%%%%%%%%%%%
\begin{eqnarray}
  |\mu|^2 &=& \frac{1}{\tan^2\beta -1} \biggl \{ \left [
    m_{H_d}^2 + {\bf m}_{L_iH_d}^2 \frac{v_i}{v_d} + \kappa_i^* \mu
    \frac{v_i}{v_d} \right ]
  - \left[ m_{H_u}^2 + |\kappa_i|^2 -\frac{1}{2}(g^2+g_2^2)v_i^2
    -D_i~\frac{v_i}{v_u} \right] \tan^2\beta \biggr \} - \frac{M_Z^2}{2},
  \label{muB1}
  \\
  B &=& \frac{\sin 2\beta}{2} \left [ 
    (m_{H_d}^2 - m_{H_u}^2 + 2|\mu|^2 + |\kappa_i|^2) 
    + ({\bf m}_{L_iH_d}^2 + \kappa_i^*\mu)\frac{v_i}{v_d}
    -D_i~\frac{v_i}{v_u} \right ],
    \label{muB2}
\end{eqnarray}
\begin{eqnarray}
  && v_1 [({\bf m}_L^2)_{11} + |\kappa_1|^2 + D'] 
  + v_2 [({\bf m}_L^2)_{21} + \kappa_1 \kappa_2^*] 
  + v_3 [({\bf m}_L^2)_{31} + \kappa_1 \kappa_3^*]
  = - [{\bf m}_{H_d L_1}^2 + \mu^*\kappa_1]v_d + D_1 v_u, \nonumber \\
  && v_1 [({\bf m}_L^2)_{12} + \kappa_2 \kappa_1^*] 
  + v_2 [({\bf m}_L^2)_{22} + |\kappa_2|^2 + D'] 
  + v_3 [({\bf m}_L^2)_{32} + \kappa_2 \kappa_3^*] 
  = - [{\bf m}_{H_d L_2}^2 + \mu^*\kappa_2]v_d + D_2 v_u,   
  \label{tadpoles} \\
  && v_1 [({\bf m}_L^2)_{13} + \kappa_3 \kappa_1^*] 
  + v_2 [({\bf m}_L^2)_{23} + \kappa_3 \kappa_2^*] 
  + v_3 [({\bf m}_L^2)_{33} + |\kappa_3|^2 + D'] 
  = - [{\bf m}_{H_d L_3}^2 + \mu^*\kappa_3]v_d + D_3 v_u, \nonumber
\end{eqnarray}
where $D'=M_Z^2 \frac{\cos2\beta}{2} + (g^2+g_2^2)
\frac{\sin^2\beta}{2}(v^2-v_u^2-v_d^2)$. The tadpole equations
(\ref{tadpoles}) can be easily solved and we get for the sneutrino vev's
\begin{equation}
  v_i~=\frac{\det W_i}{\det W}, \qquad i=1,2,3,
\end{equation}
where
\begin{equation}
  W~= \left(
    \begin{array}{lll}
      ({\bf m}_L^2)_{11} + |\kappa_1|^2 + D'   &
      ({\bf m}_L^2)_{21} + \kappa_1 \kappa_2^* & 
      ({\bf m}_L^2)_{31} + \kappa_1 \kappa_3^* \\
      ({\bf m}_L^2)_{12} + \kappa_2 \kappa_1^* &
      ({\bf m}_L^2)_{22} + |\kappa_2|^2 + D'   &
      ({\bf m}_L^2)_{32} + \kappa_2 \kappa_3^* \\
      ({\bf m}_L^2)_{13} + \kappa_3 \kappa_1^* &
      ({\bf m}_L^2)_{23} + \kappa_3 \kappa_2^* &
      ({\bf m}_L^2)_{33} + |\kappa_3|^2 + D'
    \end{array}
    \right ),
\end{equation}
and $W_i$ can be obtained from $W$ by replacing the $i$-th column with
\begin{equation}
  \left (
    \begin{array}{l}
     - [{\bf m}_{H_d L_1}^2 + \mu^*\kappa_1]v_d + D_1 v_u \\
     - [{\bf m}_{H_d L_2}^2 + \mu^*\kappa_2]v_d + D_2 v_u \\
     - [{\bf m}_{H_d L_3}^2 + \mu^*\kappa_3]v_d + D_3 v_u
    \end{array}
    \right ).
\end{equation}
%%%%%%%%%%%%%%%%%%%%
\end{widetext}
%%%%%%%%%%%%%%%%%%%%

The newly obtained $v_i$'s are incorporated in the scheme in the
following way:
\begin{eqnarray}
  v_u &=& \sin\beta \sqrt{v^2 - (v_1^2 + v_2^2 + v_3^3)}, \\
  v_d &=& \cos\beta \sqrt{v^2 - (v_1^2 + v_2^2 + v_3^3)},
\end{eqnarray}
$v^2=(246 \GeV)^2$, which preserves the definition of the $\tan\beta$
and $v^2=v_u^2+v_d^2+v_1^2+v_2^2+v_3^2$. Notice, that in this model
sneutrino vev's contribute to the mass of the $Z$ and $W$ bosons. The
equations (\ref{muB1})--(\ref{tadpoles}) are solved subsequently until
convergence and self-consistency of the results is obtained. In practice
three iterations turned out to be enough. After this procedure we add
also the dominant radiative corrections \cite{barger}, and get back to
the $m_Z$ scale to obtain the mass spectrum of the model.

\section{Neutrino magnetic moments}

It is well known that in the RpV supersymmetric models only one neutrino
gains mass after the diagonalization of the neutrino--neutralino mass
matrix. The remaining contributions come from the sneutrino vev's at the
tree level and from the one-loop processes. The main one-loop mechanism
involves decomposing a~Majorana neutrino--neutrino vertex into
a~particle--sparticle loop, which basically is either the quark--squark,
or the (charged) lepton--slepton loop. These contributions will be
proportional to some functions of the masses of the particles and $({\bf
  \Lambda}_D)^2$ or $({\bf \Lambda}_E)^2$. Detailed discussions can be
found, e.g., in Refs.~\cite{Haug,Bhatta,Abada,Grossman,mg06,mg,dl}.

Neutrinos are neutral particles and as such cannot interact with
photons. However, this interaction is possible through the one-loop
mechanism, in which charged particles appear in the loop. Therefore, an
effective neutrino magnetic moment may be generated. Due to the CPT
theorem, Majorana neutrinos cannot have diagonal magnetic moments, which
act between the same flavours of neutrinos, but they may have the
off-diagonal transition magnetic moments \cite{kayser82}. To be more
precise, the transition Majorana magnetic moment acts between $\nu_{i
  L}$ and $\nu_{j L}^c$ chiral components of Majorana neutrinos,
assuming gauge theory with only left-handed neutrinos. As a consequence
it violates the total lepton number by two units ($\Delta L =2$), and
can be discussed provided that $R$-parity violation occurs. The
effective Hamiltonian takes the form
\begin{equation}
  H^M_{\rm eff} = \frac{1}{2} \mu_{ij} \bar\nu_{iL} \sigma^{\alpha\beta}
  \nu_{j L}^c F_{\beta\alpha} + {\rm h.c.}
\label{Heff}
\end{equation}
%

%%%%%%%%%%%%%%%%%%%%%%%%%%%%%%%%%%%%%%%%%%%%%%%%%%%%%%%%%%%%%%%%%%%%%%%%
\begin{figure*}
  \centering
  \includegraphics[width=0.19\textwidth]{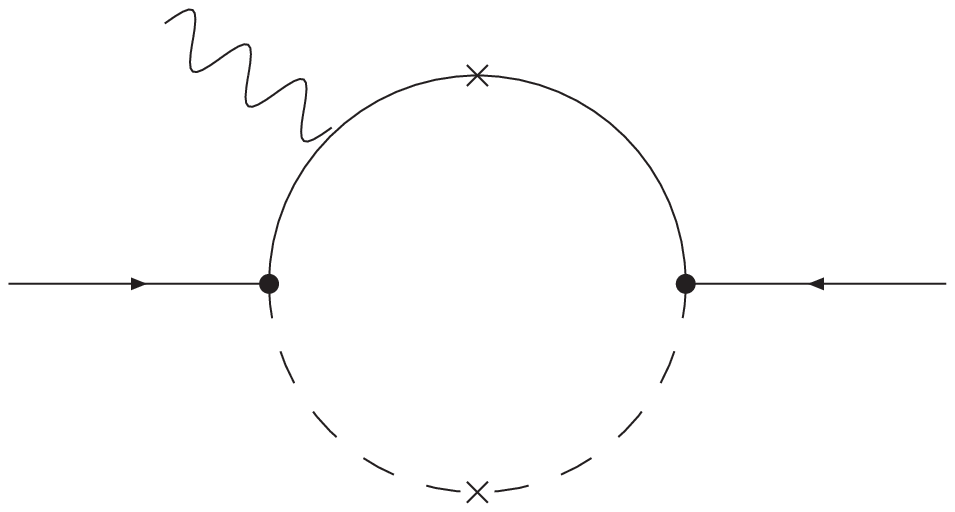} 
  \includegraphics[width=0.19\textwidth]{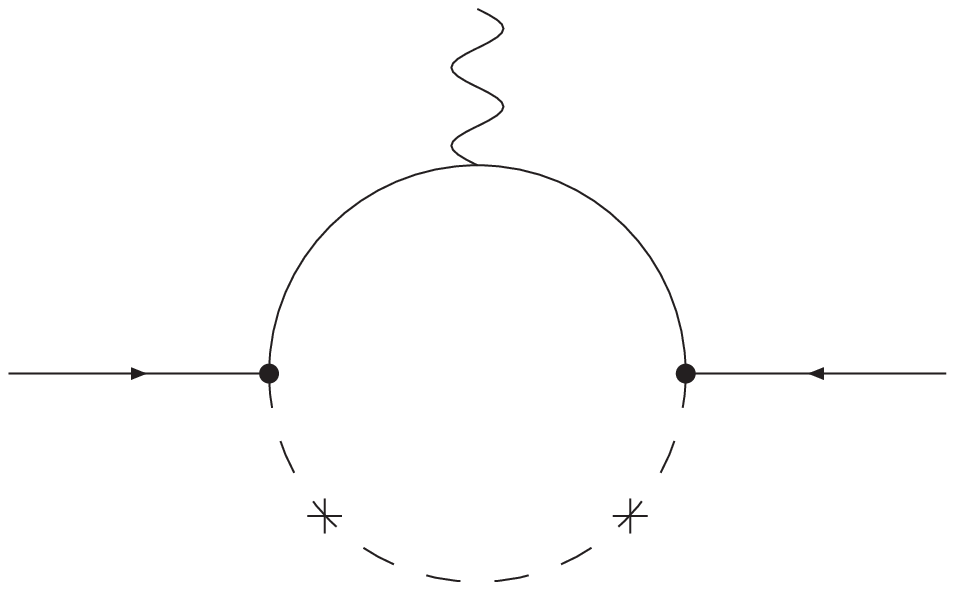}
  \includegraphics[width=0.19\textwidth]{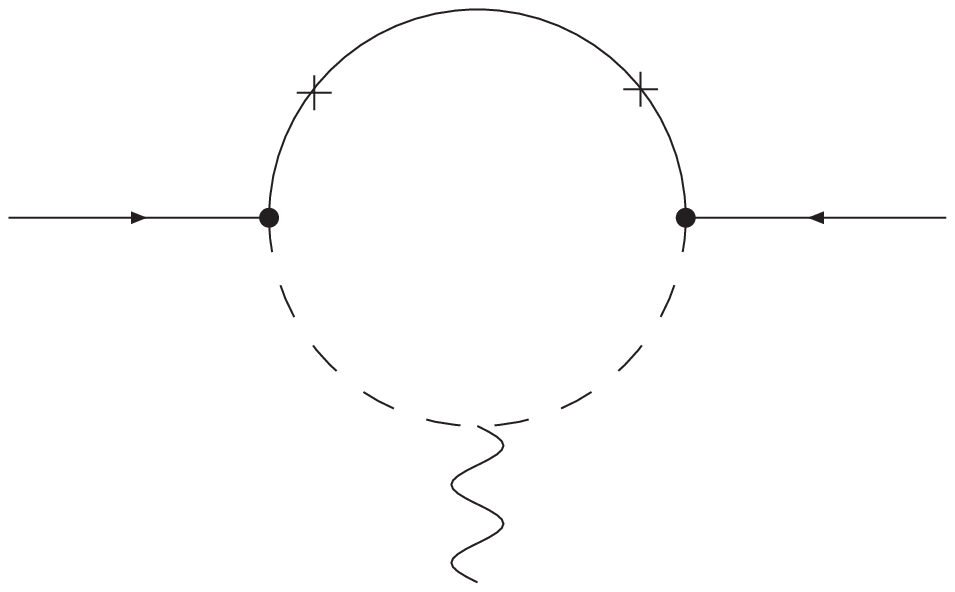} 
  \includegraphics[width=0.19\textwidth]{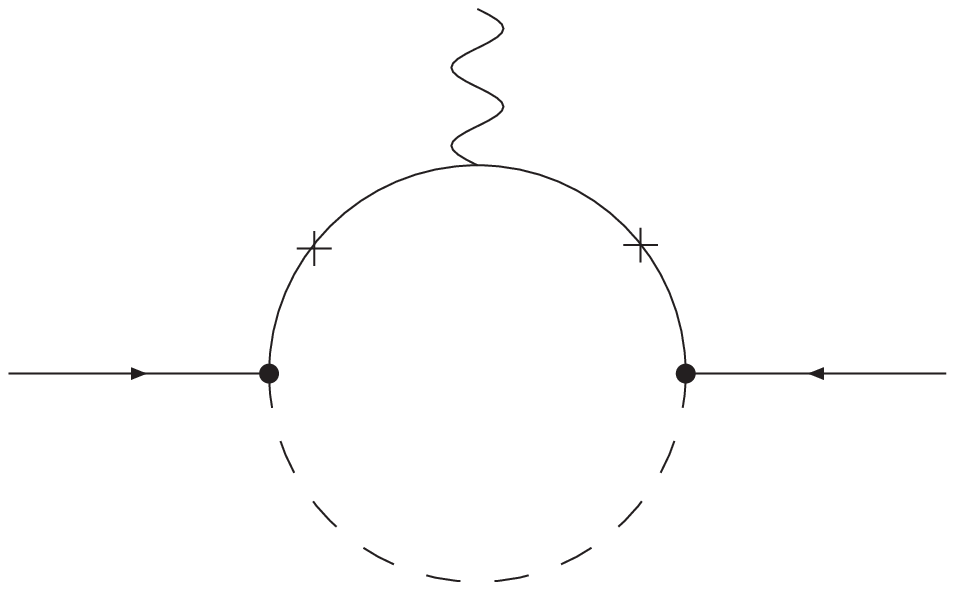} 
  \includegraphics[width=0.19\textwidth]{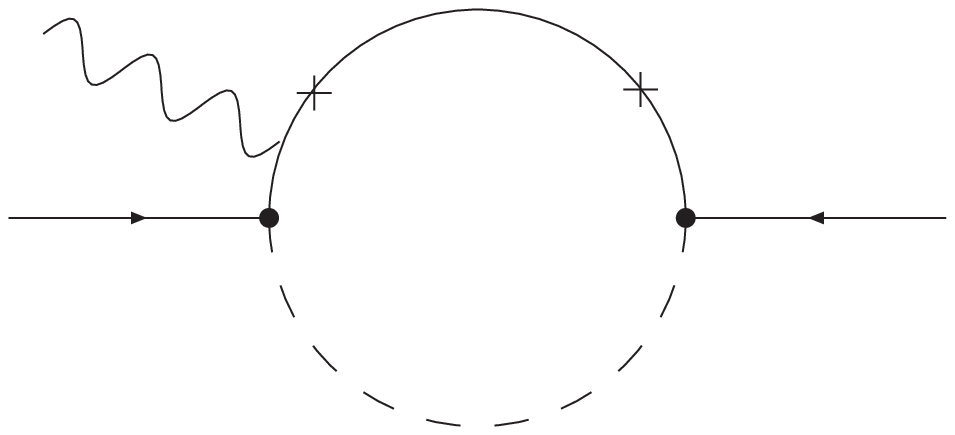}
  \centerline{\hfill(a)\hfill(b)\hfill(c)\hfill(d)\hfill(e)\hfill}
  \caption{\label{diagrams} Feynman diagrams with two mass insertions
    inside the loop contributing to the Majorana neutrino transition
    magnetic moments.}
\end{figure*}
%%%%%%%%%%%%%%%%%%%%%%%%%%%%%%%%%%%%%%%%%%%%%%%%%%%%%%%%%%%%%%%%%%%%%%%%

The contributing Feynman diagrams are presented on
Fig.~\ref{diagrams}. We will evaluate the amplitudes of these processes
in the interaction basis, in which the vertex coupling constants are
known directly from the mixing matrices. However, one has to take into
account, that the propagators of particles travelling inside the loop
must be written for their mass eigenstates. This can be done using the
mixed-propagator formalism \cite{beuthe}. Assume the following relation
between the mass ($j$) and interaction ($\alpha$) eigenstates of some
particle,
\begin{equation}
  \Phi_\alpha = \sum_j V_{\alpha j} \Phi_j.
\end{equation}
Then, the amplitude in the weak basis can be written in terms of the
Green function and expanded in the mass basis
\begin{equation}
  {\cal A}_\alpha \sim G_{\Phi_\alpha} = \sum_j V_{\alpha j} G_j V_{j \alpha}^\dagger,
\end{equation}
so the flavour changing amplitude may be written as
\begin{equation}
  {\cal A}(\alpha\to\beta) = \sum_j
  V_{\beta j} {\cal A}_j V_{j \alpha}^\dagger.
\end{equation}
A~more detailed derivation and discussion of this result can be found in
Sect.~4.2 of Ref.~\cite{beuthe}. 

The matrix $V$ is obtained during the diagonalization procedure of the
mixing matrices and its columns are the eigenvectors of the mixing
matrix. The magnetic moment will consist of the product of numerical
constants, coupling constants, and the function $\cal I$ describing the
loop integral, so it may be written in a~general form as
\begin{equation}
\mu_\nu = \left[\sum N_c\ C_1 C_2 X_1 X_2\ Q\ {\cal I}\right] 2m_e\ \mu_B,
\end{equation}
where $N_c$ is the color index (=3 for quarks and squarks, =1
otherwise), $C_{1,2}$ are the trilinear coupling constants, $X_{1,2}$
are the mass insertions residing inside the loop, $Q$ is the charge of
the particle the photon is attached to (in units of the elementary
charge), $m_e$ is the electron mass ($2m_e\approx 10^{-3}\GeV$), and
$\mu_B$ the Bohr magneton. The sum runs over all mass eigenstates which
form every mixed particle present in the loop.

The masses and couplings in our approach are calculated from the
boundary conditions at the $m_Z$ and $m_{\rm GUT}$ scales (see
Sect.~III). The loop integrals ${\cal I}$ can be derived analytically
using standard techniques of integration in the Minkowski
space. Denoting by $f_1,f_2,\dots$ and $b_1,b_2,\dots$ the fermions and
bosons which appear in the loop, and by $m_{f,b}$ their respective
masses, the loop integrals for diagrams (a)--(e) depicted on
Fig.~\ref{diagrams} read
%
%%%%%%%%%%%%%%%%%%%%
\begin{widetext}
%%%%%%%%%%%%%%%%%%%%
\begin{eqnarray}
  {\cal I}^{(a)} 
  &=& m_{f_1} m_{f_2}\ {\cal F}_3(f_1,f_2,b_1,b_2) 
  + {\cal F}_4(f_1,f_2,b_1,b_2)
  + m_{f_1} m_{f_2}\ {\cal F}_3(f_2,f_1,b_1,b_2) 
  + {\cal F}_4(f_2,f_1,b_1,b_2), \label{Ia}\\
  {\cal I}^{(b)} 
  &=& m_{f_1} \ {\cal F}_3(f_1,b_1,b_2,b_3), \label{Ib}\\
  {\cal I}^{(c)} 
  &=& 4\ {\cal F}_4(b_1,f_1,f_2,f_3), \label{Ic}\\
  {\cal I}^{(d)} 
  &=& m_{f_1} m_{f_2} (m_{f_2} 
  + m_{f_3})\ {\cal F}_3(f_2,f_1,f_3,b_1)
  + (2m_{f_1} + 3m_{f_2} + m_{f_3})\ {\cal F}_4(f_2,f_1,f_3,b_1), \label{Id}\\
  {\cal I}^{(e)} 
  &=& m_{f_1} (m_{f_1}m_{f_2}+m_{f_2}m_{f_3}+m_{f_3}m_{f_1})
  {\cal F}_3(f_1,f_2,f_3,b_1)
   + (5m_{f_1} + 2m_{f_2} + 2m_{f_3})\ {\cal F}_4(f_1,f_2,f_3,b_1) \nonumber \\
  &+& m_{f_3} (m_{f_1}m_{f_2}+m_{f_2}m_{f_3}+m_{f_3}m_{f_1})
  {\cal F}_3(f_3,f_1,f_2,b_1)
  + (5m_{f_3} + 2m_{f_2} + 2m_{f_1})\ {\cal F}_4(f_3,f_1,f_2,b_1),
  \label{Ie}
\end{eqnarray}
where the functions ${\cal F}_{3,4}$ are given by
\begin{eqnarray}
  (16\pi^2) {\cal F}_3(a,b,c,d) 
  &=& (16\pi^2) \int_{-\infty}^{\infty} \frac{d^4k}{(2\pi)^4}
  \frac{1}{(k^2-m_a^2)^2 (k^2 - m_b^2) 
    (k^2-m_c^2) (k^2 - m_d^2)} \nonumber \\
  &=& \frac{m_a^2 \log(m_a^2) \left(
      \frac{1}{m_a^2-m_b^2}+\frac{1}{m_a^2-m_c^2}+\frac{1}{m_a^2-m_d^2}
      -1 \right) -1}{(m_a^2-m_b^2)(m_a^2-m_c^2)(m_a^2-m_d^2)}
  - \frac{m_b^2 \log(m_b^2)}{(m_b^2-m_a^2)(m_b^2-m_c^2)(m_b^2-m_d^2)} 
  \nonumber \\
  &-& \frac{m_c^2 \log(m_c^2)}{(m_c^2-m_a^2)(m_c^2-m_b^2)(m_c^2-m_d^2)}
   -  \frac{m_d^2 \log(m_d^2)}{(m_d^2-m_a^2)(m_d^2-m_b^2)(m_d^2-m_c^2)},
  \\
  (16\pi^2) {\cal F}_4(a,b,c,d) 
  &=& (16\pi^2) \int_{-\infty}^{\infty} \frac{d^4k}{(2\pi)^4}
  \frac{k^2}{(k^2-m_a^2)^2 (k^2 - m_b^2) 
    (k^2-m_c^2) (k^2 - m_d^2)} \nonumber \\
  &=& \frac{ m_a^4 \log(m_a^2) \left(
      \frac{1}{m_a^2-m_b^2}+\frac{1}{m_a^2-m_c^2}+\frac{1}{m_a^2-m_d^2}
      -2 \right) -1}{(m_a^2-m_b^2)(m_a^2-m_c^2)(m_a^2-m_d^2)} 
  - \frac{m_b^4 \log(m_b^2)}{(m_b^2-m_a^2)(m_b^2-m_c^2)(m_b^2-m_d^2)}
  \nonumber \\
  &-& \frac{m_c^4 \log(m_c^2)}{(m_c^2-m_a^2)(m_c^2-m_b^2)(m_c^2-m_d^2)}
  -   \frac{m_d^4 \log(m_d^2)}{(m_d^2-m_a^2)(m_d^2-m_b^2)(m_d^2-m_c^2)}.
\end{eqnarray}
%%%%%%%%%%%%%%%%%%%%
\end{widetext}
%%%%%%%%%%%%%%%%%%%%

We have kept denotations from Ref.~\cite{mg17}. Please notice also
that a~typo has been corrected in Eq.~(\ref{Ib}), and new terms appeared
in Eqs.~(\ref{Ia}) and (\ref{Ie}). 

\section{Results}

In order to contribute to the neutrino magnetic moment, the loop must
contain charged particles. The quarks--squarks and leptons--sleptons
have already been discussed elsewhere (see Sect. I). The discussion on
the bilinear insertions on the neutrino lines, possible due to the
neutrino--neutralino mixing, can be found in Refs.~\cite{mg18,mg19}. In
this work we have focused on the charged Higgs--sleptons and the
leptons--charginos interaction eigenstates.

Altogether we have identified 45 diagrams of the forms given on
Fig.~\ref{diagrams} containing the particles in question. The way of
obtaining them is a~straightforward but rather tedious task, and we have
used a~computer program to match all possible vertices and internal
lines to the given patterns. The trilinear vertices can be constructed
directly from the superpotential (\ref{W}), and the possible bilinear
mass insertions are defined by the mixing matrices, as given in
Ref.~\cite{mSUGRA}.

A~few conditions must have been met for the result to be accepted. First
of all, during the RGE evolution some of the particles may get negative
mass parameters squared. Such tachyonic behaviour resulted in immediate
rejecting of the given input parameters. Also, finite values for the
Yukawa couplings, which tend to `explode' for too low or too high values
of the $\tan\beta$, were required. When analyzing the results, we have
rejected points, for which the masses of the superparticles where too
low. The mass limits we have used where taken from the Particle Data
Group report \cite{pdg2010}, and they read: $m_{\tilde\chi^0_1} > 46
\GeV$, $m_{\tilde\chi^0_2} > 62 \GeV$, $m_{\tilde\chi^0_3} > 100 \GeV$,
$m_{\tilde\chi^0_4} > 116 \GeV$, $m_{\tilde\chi^\pm_1} > 94 \GeV$,
$m_{\tilde\chi^\pm_2} > 94 \GeV$, $m_{\tilde e} > 107 \GeV$,
$m_{\tilde\mu} > 94 \GeV$, $m_{\tilde\tau} > 82 \GeV$, $m_{\tilde q} >
379 \GeV$, $m_{\tilde g} > 308 \GeV$. This is by far the conservative
choice of constraints. The new results from the LHC projects push the
quoted above limits to higher values, closer to the 1~TeV range. We do
not use them here for various reasons. Firstly, most of the results are
still marked as preliminary, and more careful data analysis is being
performed right now. Also, the LHC data are interpreted within the
simplest SUSY models, with many additional assumptions, like the absence
of $R$--parity violation, and equality of the gluino and squark masses.
Theses assumptions are not true in the model used here.
 
\subsection{Constraining the model's parameter space with the magnetic
  moments}

\begin{figure*}
  \centering
  \includegraphics[width=0.85\linewidth]{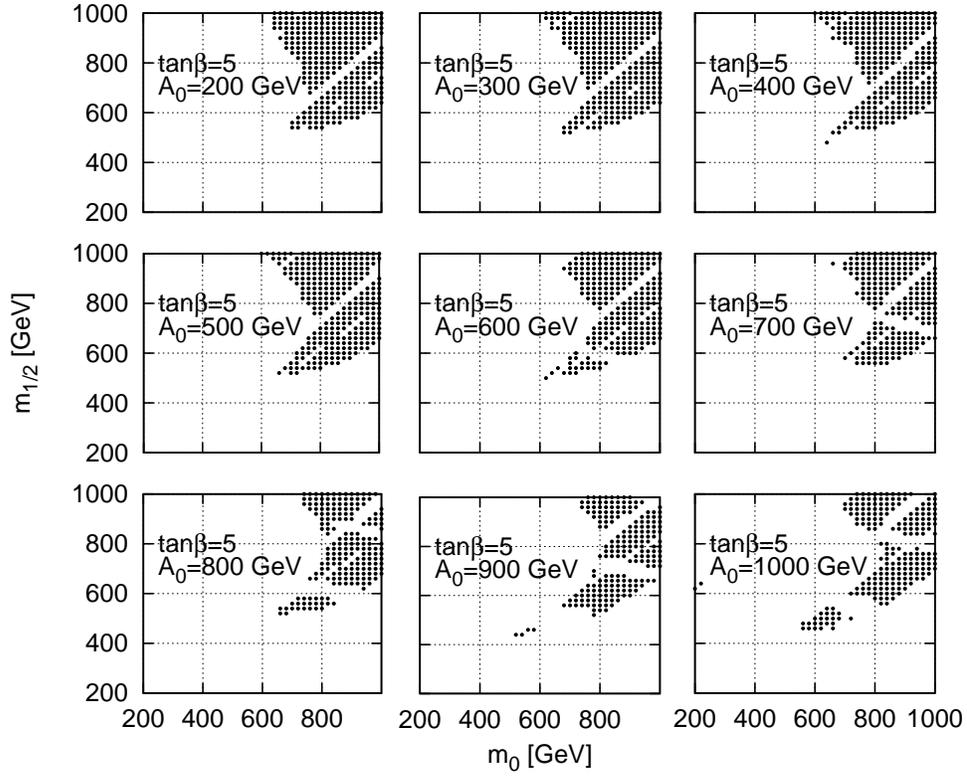}
  \caption{Allowed parameter space of the model for $\tan\beta=5$ and
    $\mu>0$. The parameter $A_0$ changes as indicated on the panels.}
  \label{fig:tgb05}
\end{figure*}

\begin{figure*}
  \centering
  \includegraphics[width=0.85\linewidth]{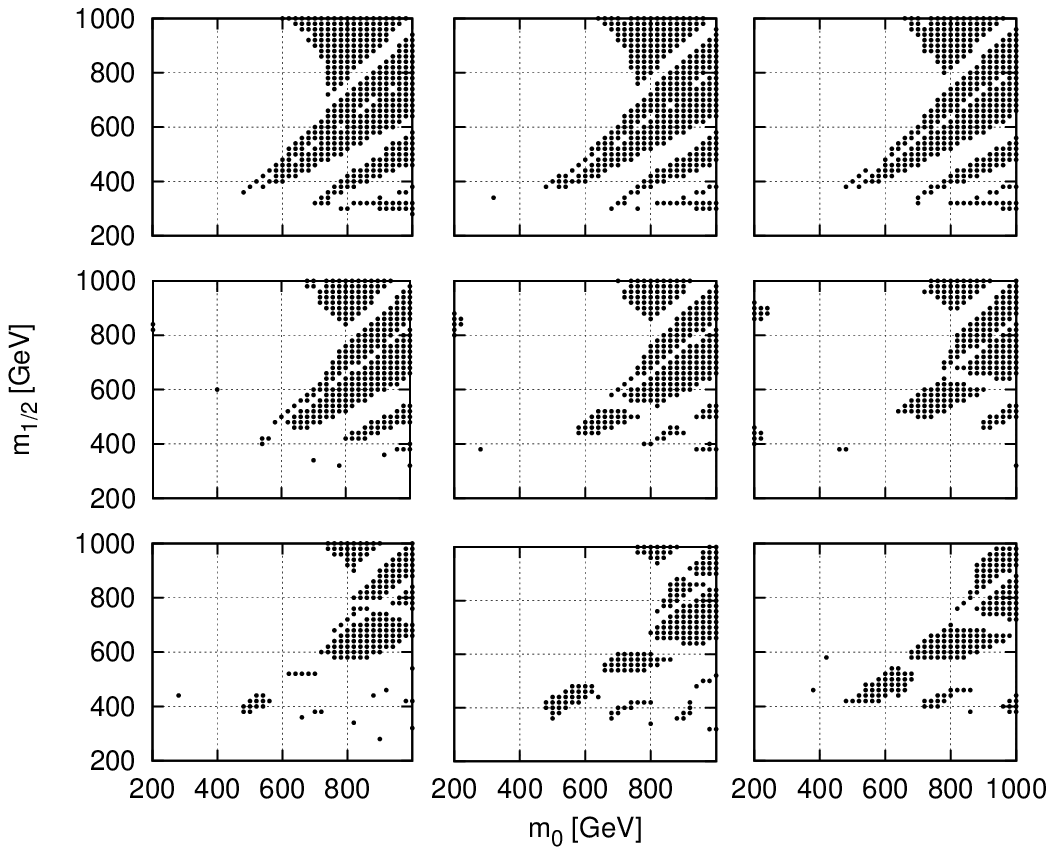}
  \caption{Like Fig.~\ref{fig:tgb05} but for $\tan\beta=10$.}
  \label{fig:tgb10}
\end{figure*}

\begin{figure*}
  \centering
  \includegraphics[width=0.85\linewidth]{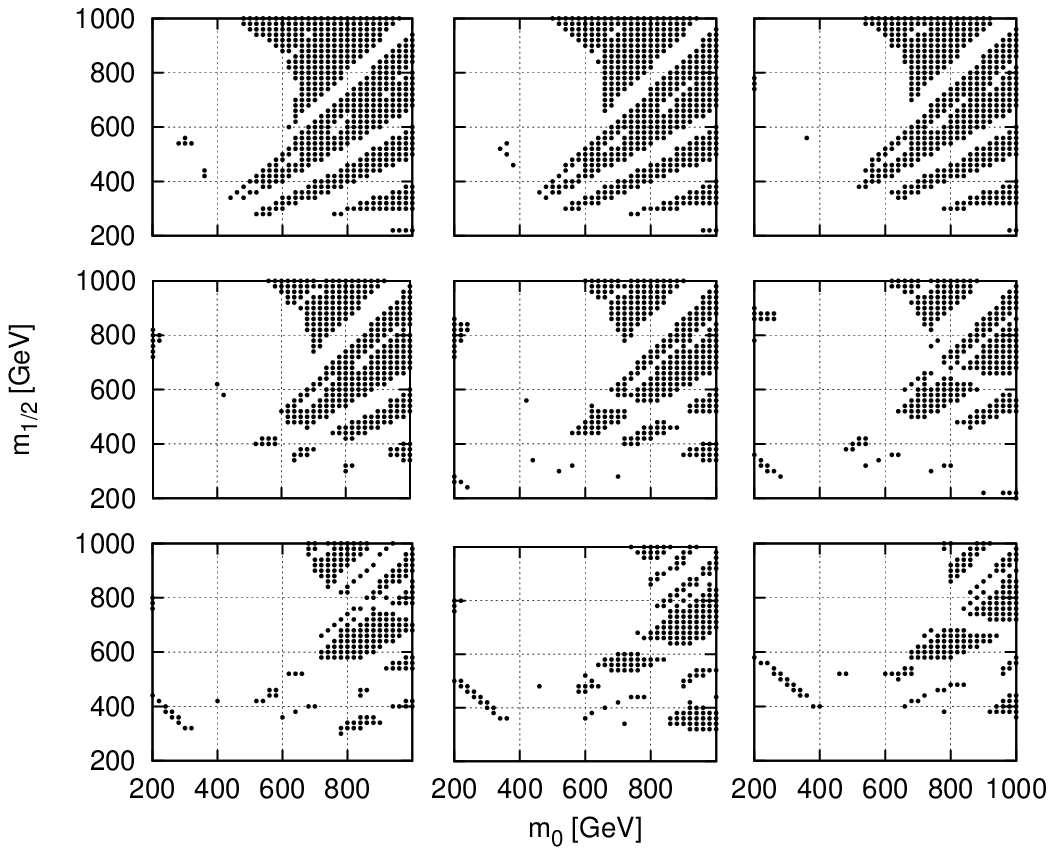}
  \caption{Like Fig.~\ref{fig:tgb05} but for $\tan\beta=15$.}
  \label{fig:tgb15}
\end{figure*}

\begin{figure*}
  \centering
  \includegraphics[width=0.85\linewidth]{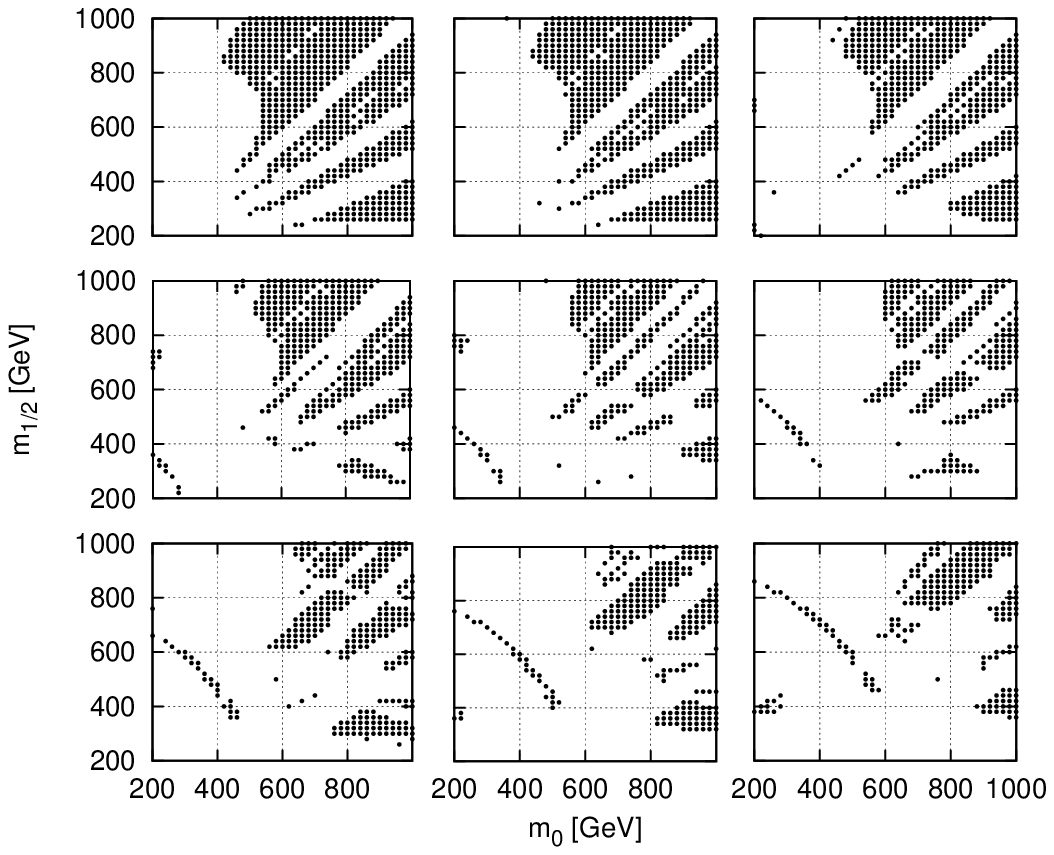}
  \caption{Like Fig.~\ref{fig:tgb05} but for $\tan\beta=20$.}
  \label{fig:tgb20}
\end{figure*}

\begin{figure*}
  \centering
  \includegraphics[width=0.85\linewidth]{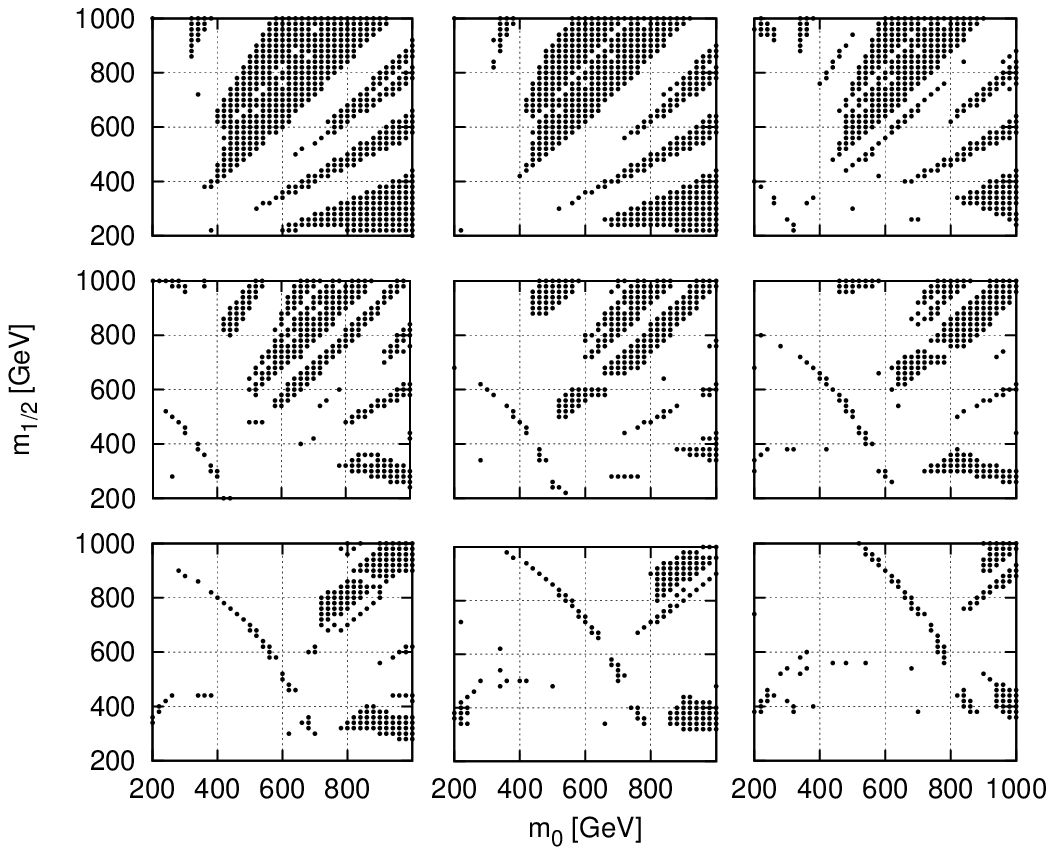}
  \caption{Like Fig.~\ref{fig:tgb05} but for $\tan\beta=25$.}
  \label{fig:tgb25}
\end{figure*}

\begin{figure*}
  \centering
  \includegraphics[width=0.85\linewidth]{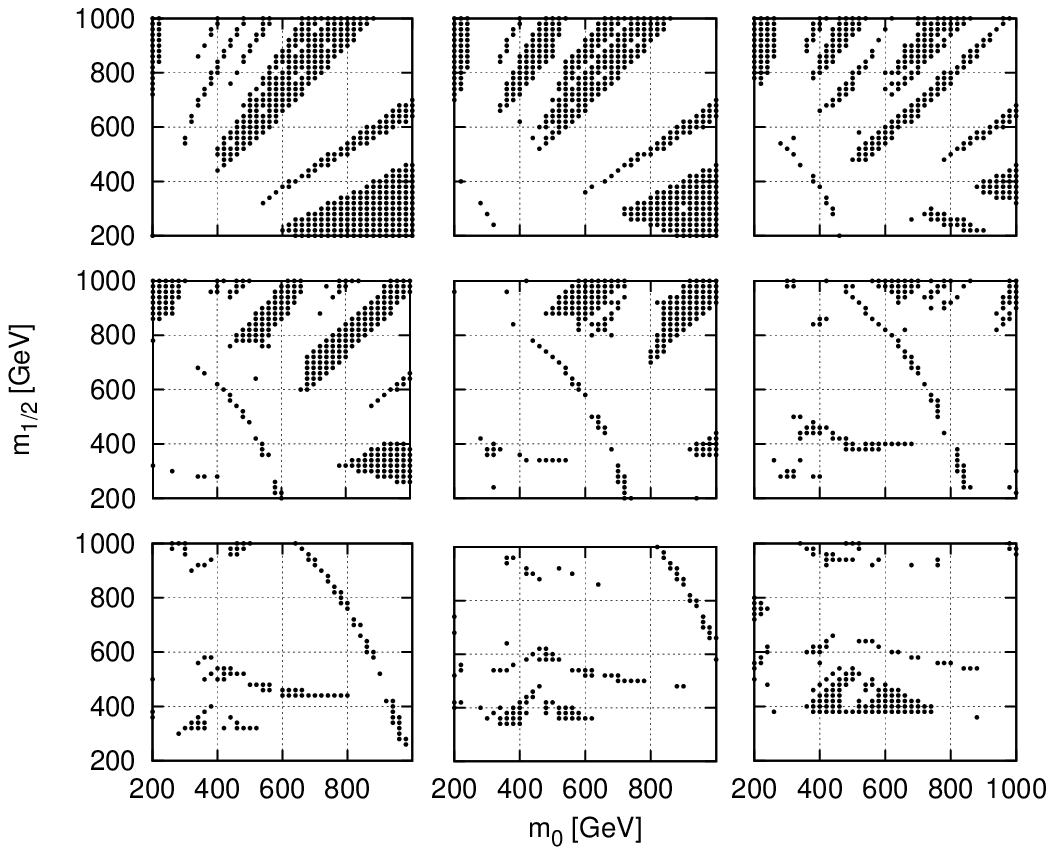}
  \caption{Like Fig.~\ref{fig:tgb05} but for $\tan\beta=30$.}
  \label{fig:tgb30}
\end{figure*}

\begin{figure*}
  \centering
  \includegraphics[width=0.85\linewidth]{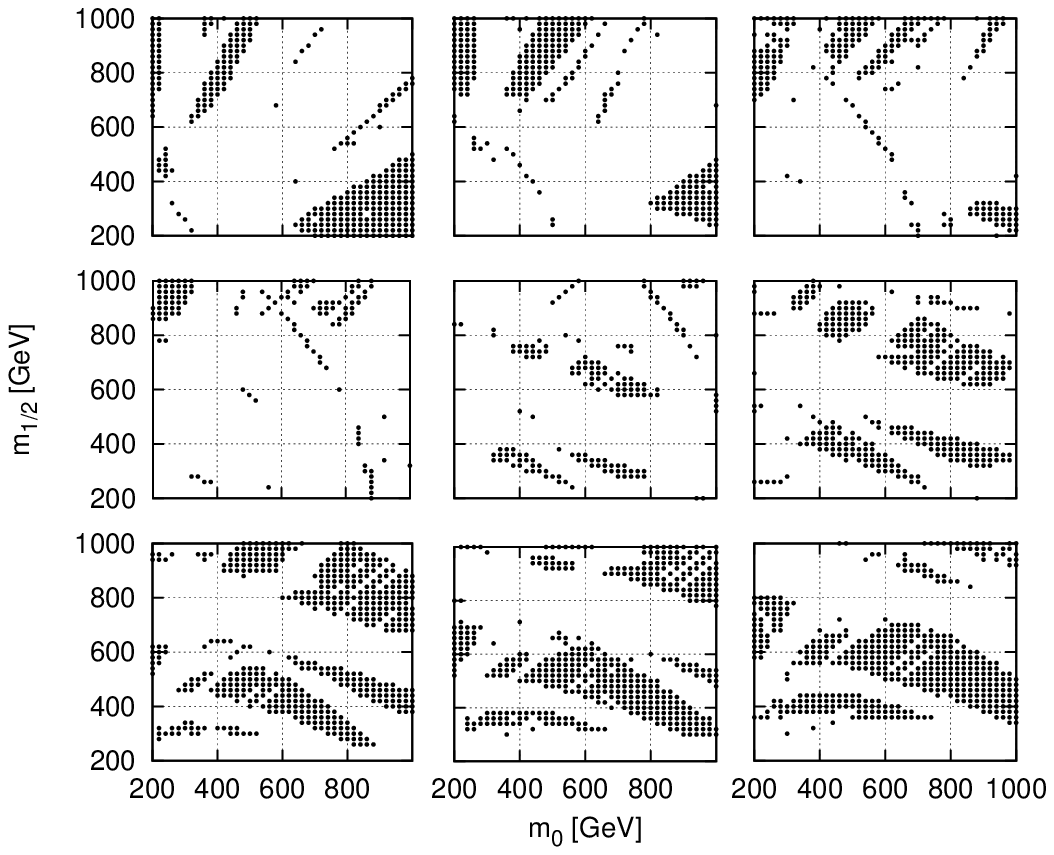}
  \caption{Like Fig.~\ref{fig:tgb05} but for $\tan\beta=35$.}
  \label{fig:tgb35}
\end{figure*}

\begin{figure*}
  \centering
  \includegraphics[width=0.85\linewidth]{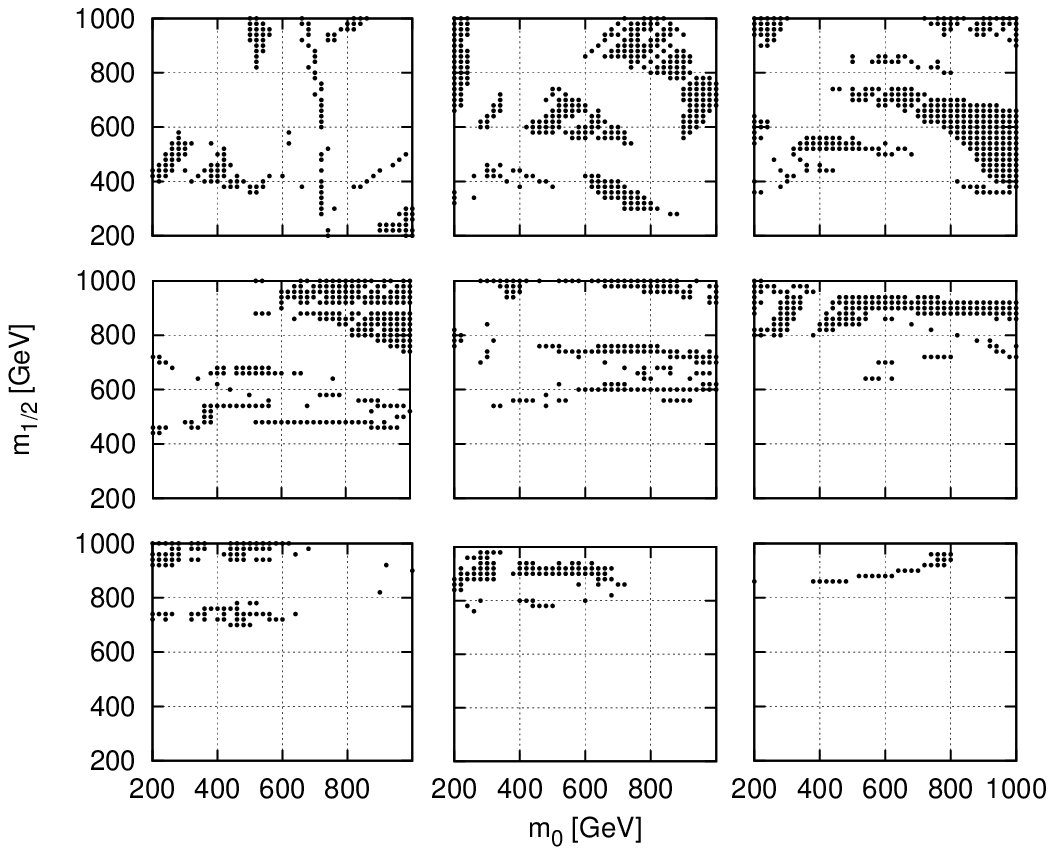}
  \caption{Like Fig.~\ref{fig:tgb05} but for $\tan\beta=40$. Please
    notice, that for such high value the model breaks down and the
    results are no longer reliable.}
  \label{fig:tgb40}
\end{figure*}

There are different assessments and constraints put on the neutrino
magnetic moment. Analyzing the impact of the solar neutrino data on the
neutrino spin-flavour-precession mechanism, the authors of
Ref.~\cite{magmom:valle} have found a~model-dependent upper value for
the magnetic moment to be few$\times 10^{-12}\mu_B$. On the other hand,
direct searches in the MUNU experiment resulted in the upper limit to be
$9\times 10^{-11}\mu_B$ \cite{magmom:MUNU}. In this paper we adopt
a~conservative limit and reject points, for which $\mu_\nu \ge
10^{-10}\mu_B$. A~more strict approach would result in a~substantially
narrower parameter space of the model.

The results are presented on Figs.~\ref{fig:tgb05}--\ref{fig:tgb40}, for
$\mu>0$, $\Lambda_0=10^{-4}$, $A_0=200-1000 \GeV$ in steps of 100,
$m_{0,1/2}=200-1000 \GeV$ in steps of 20, and $\tan\beta=5-40$ in steps
of 5. Each point represents a~valid set of input parameters, which
results in a~physically accepted low energy spectrum. Each figure
represents a~fixed $\tan\beta$, and panels read in rows from left to
right correspond to $A_0$ taking values 200 GeV, 300 GeV, \dots, 1000 GeV.

In general, there is no global regularity in these results. Certain
regions are excluded due to unaccepted values of the masses. The
excluded stripes, visible in several of the panels, come from too high
values of $\mu_\nu$. Starting from $\tan\beta=15$, a~parabolic-like
shape, whose position depends on the $A_0$ parameter, appears. It starts
to be visible for $A_0=400 \GeV$ and crosses the $m_0-m_{1/2}$ plane
with increasing $A_0$. [We are not sure about the origin of this
`stability line'. As a~side remark we add, that it appeared also in our
analysis of the masses of the lightest Higgs bosons $h^0$ in the context
of the CDF-D0 discovery \cite{Tevatron}. Constructing similar plots, but
using the condition, that $120\GeV < m_{h^0} < 140\GeV$, we have
observed very similar parabolas.] We notice also, that for
$\tan\beta=40$ the points look totally random, which shows that the
model breaks down in this region due to too high values of the Yukawa
couplings.

\subsection{$\Lambda_0$ dependence}

\begin{figure}
  \centering
  \includegraphics[width=0.4\textwidth]{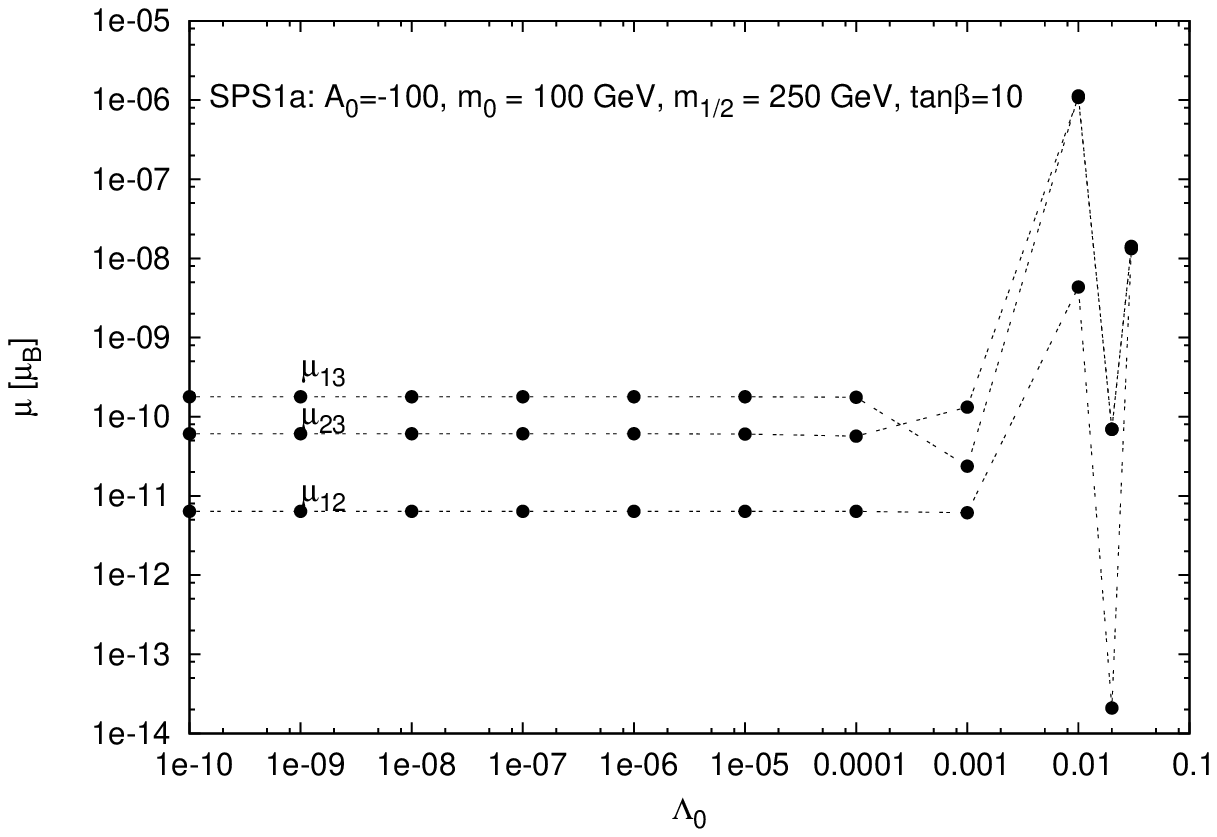}
  \caption{Contributions to the magnetic moment of the neutrino for
    SPS1a SUGRA point.}
  \label{fig:sps1a}
\end{figure}

\begin{figure}
  \centering
  \includegraphics[width=0.4\textwidth]{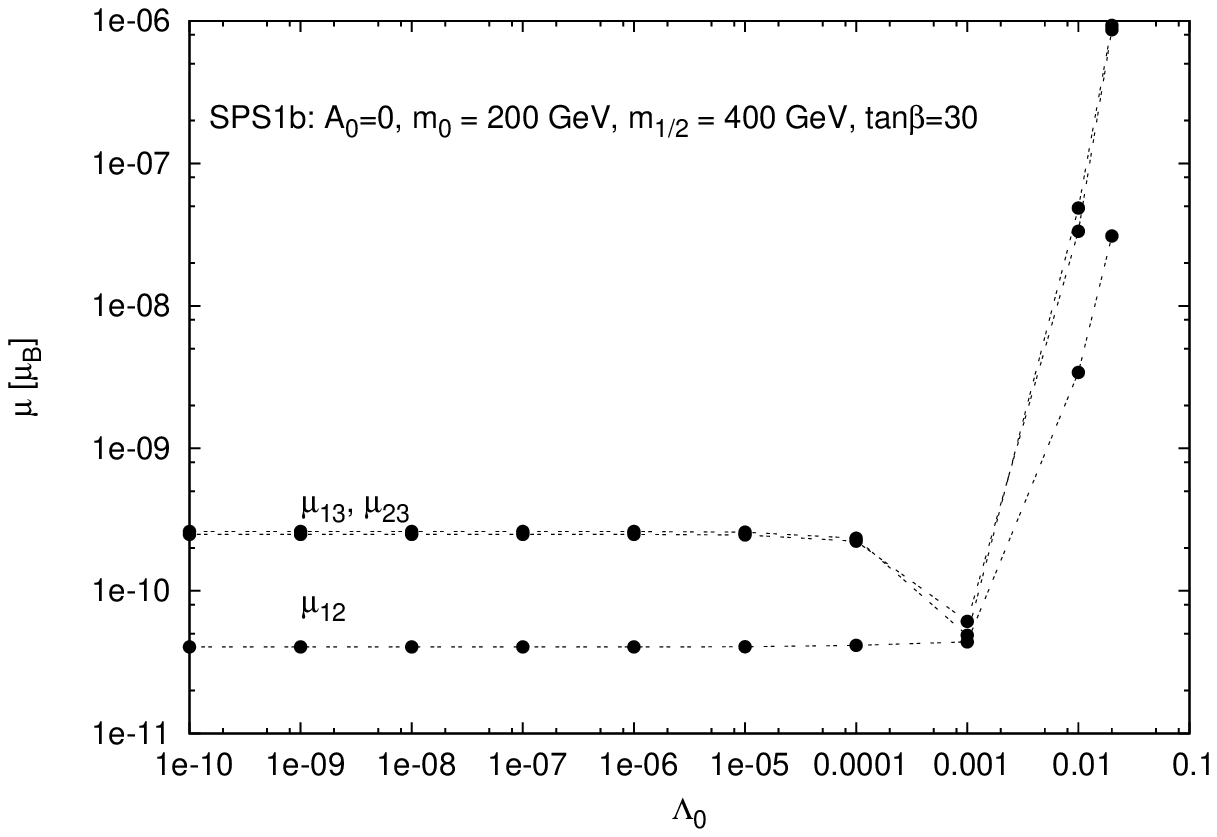}
  \caption{Contributions to the magnetic moment of the neutrino for
    SPS1b SUGRA point.}
  \label{fig:sps1b}
\end{figure}

\begin{figure}
  \centering
  \includegraphics[width=0.4\textwidth]{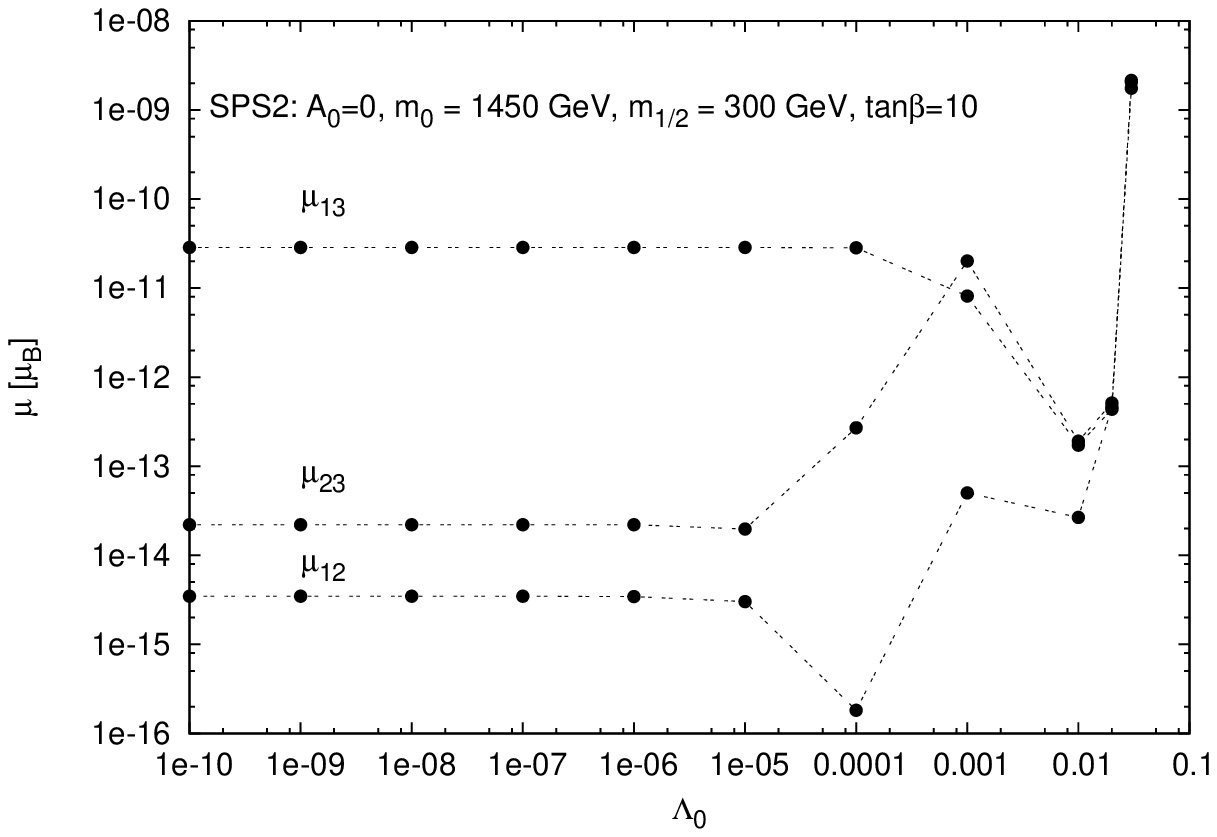}
  \caption{Contributions to the magnetic moment of the neutrino for SPS2
    SUGRA point.}
  \label{fig:sps2}
\end{figure}

\begin{figure}
  \centering
  \includegraphics[width=0.4\textwidth]{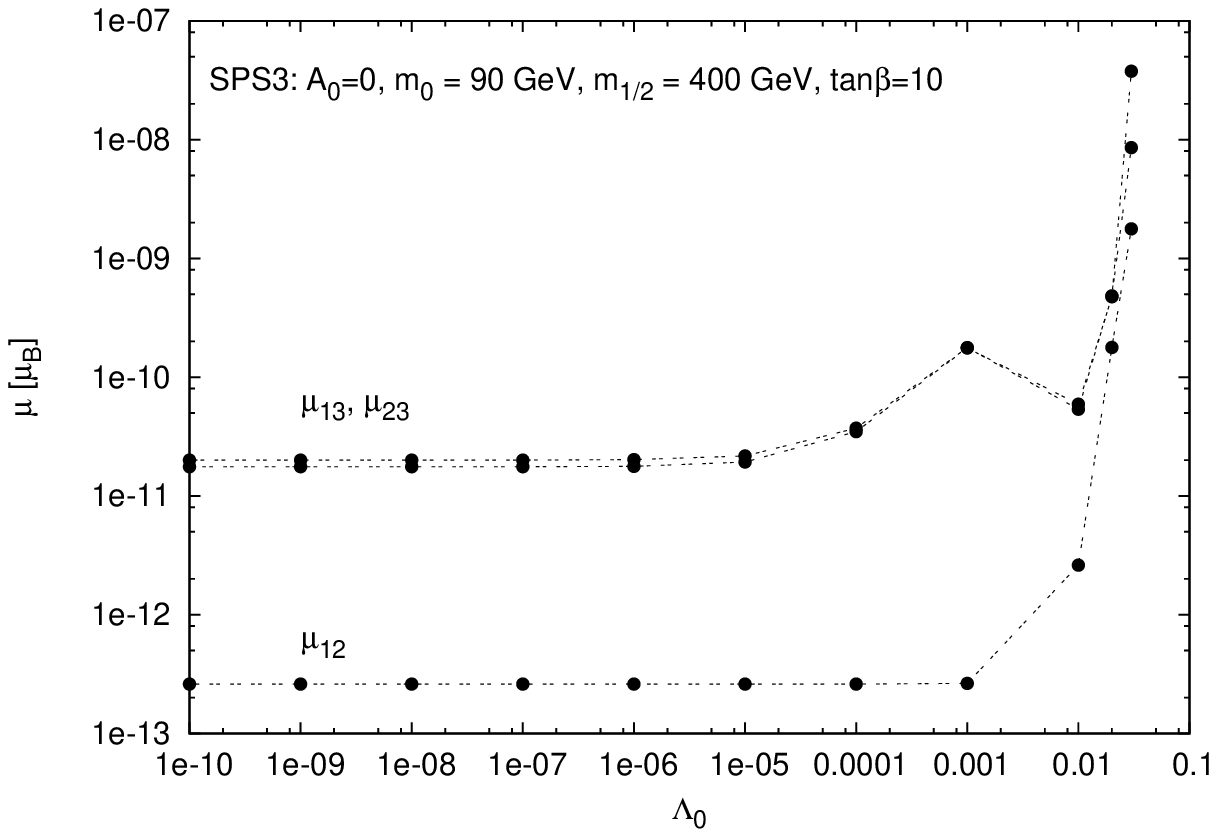}
  \caption{Contributions to the magnetic moment of the neutrino for SPS3
    SUGRA point.}
  \label{fig:sps3}
\end{figure}

Another interesting problem is how the initial value of the RpV
couplings at $m_Z$, represented by a~common parameter $\Lambda_0$,
influences the results. We have chosen to check this relation for a~few
specific sets of parameters. For this, the Snowmass SUSY benchmark
points \cite{snowmass} were used. The dependence of the resulting
contribution to the transition magnetic moments have been calculated for
the SUGRA benchmark points SPS1a, SPS1b, SPS2, and SPS3, are shown on
Figs.~\ref{fig:sps1a}--\ref{fig:sps3}. In this calculation we have not
discriminated the results which were exceeding the upper limit for
$\mu_\nu$.

The existing limits on $\bf \Lambda$'s, obtained withtin SUSY models,
oscillate around $10^{-2}-10^{-5}$, depending on the method used
\cite{lambda-limits,mg06}. We check here the range between few$\times
10^{-2}$ and $10^{-10}$. We notice that the RGE running decreases the
values of $\Lambda$'s for higher energies \cite{mg01}, thus the
parameter $\Lambda_0$ sets the upper limit on them.

We first notice, that below roughly $\Lambda_0=10^{-5}$ (in some cases
even more) the results stabilize and do not change with decreasing
$\Lambda_0$, which makes them almost indistinguishable with the
situation when all RpV couplings are set to zero. On the other hand, the
calculations broke down for $\Lambda_0 > {\rm few} \times 10^{-2}$. This
leaves a~rather narrow region of $\Lambda_0$, between 1 and 3 orders of
magnitude and only modestly exceeding $10^{-2}$, for which the RpV
couplings play a~role.

We see also that for small $\Lambda_0$ the general alignment of the
$\mu$'s resembles hierarchical structure, while for high $\Lambda_0$
this hierarchy vanishes, and the results show no regular pattern. Too
high values of $\Lambda_0$ can boost them to values close to
$10^{-6}\mu_B$, which are excluded by experiments like MUNU
\cite{magmom:MUNU} (recall upper limit from the MUNU $\sim
10^{-10}\mu_B$). Curiously, for small (or zero) values of $\Lambda_0$,
all the transition magnetic moments tend to be of the order of
$10^{-10}\mu_B$ (SPS1a, SPS1b) or lower (SPS2, SPS3).

Let us also compare the newly computed contributions with similar
contributions coming from simplest loops with no mass insertions
(trilinear RpV couplings only) and loops with bilinear insertions on the
external neutrino lines (neutrino--neutralino mixing). We do it for two
points, for which earlier calculations have been presented in
Refs.~\cite{mg18,mg09} -- the unification with {\sc low} values:
$A_0=100 \GeV$, $m_0=m_{1/2}=150$, and {\sc high} values: $A_0=500
\GeV$, $m_0=m_{1/2}=1000$. In both of these cases $\tan\beta=19$ and
$\mu>0$.
%
%%%%%%%%%%%%%%%%%%%%%%%%%%%%%%%%%%%%%%%%%%%%%%%%%%%%%%%%%%%%%%%%%%%%%%%%
%%%%%%%%%%%%%%%%%%%%%%%%%%%%%%%%%%%%%%%%%%%%%%%%%%%%%%%%%%%%%%%%%%%%%%%%
%%%%%%%%%%%%%%%%%%%%%%%%%%%%%%%%%%%%%%%%%%%%%%%%%%%%%%%%%%%%%%%%%%%%%%%%
\begin{table}[t]
  \centering
  \caption{\label{tab:low} Comparison of the magnitudes of different contributions to
    the Majorana neutrino transition magnetic moments for the input
    parameters set {\sc low} (see text). Here subscripts $1,2,3=e,\mu,\tau$.}
  \begin{tabular}{cccc} 
    \hline \hline
    $\Lambda_0$ & $\mu_{\nu_{12}}$ & $\mu_{\nu_{13}}$ & $\mu_{\nu_{23}}$ \\
    \hline
    $3.0\times 10^{-2}  $&$ 6.69\times 10^{-9}  $&$ 3.21\times 10^{-7} $&$ 4.07\times 10^{-7}$ \\
    $2.0\times 10^{-2}  $&$ 9.43\times 10^{-8}  $&$ 1.41\times 10^{-7} $&$ 1.25\times 10^{-7}$ \\
    $1.0\times 10^{-2}  $&$ 3.34\times 10^{-10} $&$ 2.40\times 10^{-9} $&$ 3.15\times 10^{-9}$ \\
    $1.0\times 10^{-3}  $&$ 5.57\times 10^{-10} $&$ 2.09\times 10^{-7} $&$ 2.12\times 10^{-7}$ \\
    $1.0\times 10^{-4}  $&$ 5.45\times 10^{-10} $&$ 4.97\times 10^{-8} $&$ 4.60\times 10^{-8}$ \\
    $1.0\times 10^{-5}  $&$ 5.42\times 10^{-10} $&$ 4.61\times 10^{-8} $&$ 4.24\times 10^{-8}$ \\
    $1.0\times 10^{-6}  $&$ 5.42\times 10^{-10} $&$ 4.58\times 10^{-8} $&$ 4.20\times 10^{-8}$ \\
    $1.0\times 10^{-7}  $&$ 5.42\times 10^{-10} $&$ 4.58\times 10^{-8} $&$ 4.20\times 10^{-8}$ \\
    $1.0\times 10^{-8}  $&$ 5.42\times 10^{-10} $&$ 4.58\times 10^{-8} $&$ 4.20\times 10^{-8}$ \\
    $1.0\times 10^{-9}  $&$ 5.42\times 10^{-10} $&$ 4.58\times 10^{-8} $&$ 4.20\times 10^{-8}$ \\
    $1.0\times 10^{-10} $&$ 5.42\times 10^{-10} $&$ 4.58\times 10^{-8} $&$ 4.20\times 10^{-8}$ \\
    \hline
    trilinear & $10^{-19...-15}$ & $10^{-19...-15}$ & $10^{-18...-15}$ \\
    bilinear  & $10^{-21...-17}$ & $10^{-21...-17}$ & $10^{-19...-17}$ \\
    \hline\hline
  \end{tabular}
\end{table}
%%%%%%%%%%%%%%%%%%%%%%%%%%%%%%%%%%%%%%%%%%%%%%%%%%%%%%%%%%%%%%%%%%%%%%%%
%%%%%%%%%%%%%%%%%%%%%%%%%%%%%%%%%%%%%%%%%%%%%%%%%%%%%%%%%%%%%%%%%%%%%%%%

%%%%%%%%%%%%%%%%%%%%%%%%%%%%%%%%%%%%%%%%%%%%%%%%%%%%%%%%%%%%%%%%%%%%%%%%
\begin{table}[t]
  \centering
  \caption{\label{tab:high} Comparison of the magnitudes of different contributions to
    the Majorana neutrino transition magnetic moments for the input
    parameters set {\sc high} (see text). Here subscripts $1,2,3=e,\mu,\tau$.}
  \begin{tabular}{cccc} 
    \hline \hline
    $\Lambda_0$ & $\mu_{\nu_{12}}$ & $\mu_{\nu_{13}}$ & $\mu_{\nu_{23}}$ \\
    \hline
    $3.0\times 10^{-2}  $&$ 6.62\times 10^{-12} $&$ 2.66\times 10^{-9} $&$ 3.07\times 10^{-9}$ \\
    $2.0\times 10^{-2}  $&$ 2.97\times 10^{-10} $&$ 1.34\times 10^{-10} $&$ 1.60\times 10^{-10}$ \\
    $1.0\times 10^{-2}  $&$ 4.89\times 10^{-11} $&$ 2.19\times 10^{-10} $&$ 2.03\times 10^{-10}$ \\
    $1.0\times 10^{-3}  $&$ 1.01\times 10^{-11} $&$ 3.75\times 10^{-7} $&$ 3.83\times 10^{-7}$ \\
    $1.0\times 10^{-4}  $&$ 2.44\times 10^{-14} $&$ 8.02\times 10^{-9} $&$ 8.09\times 10^{-9}$ \\
    $1.0\times 10^{-5}  $&$ 6.46\times 10^{-15} $&$ 4.32\times 10^{-9} $&$ 4.33\times 10^{-9}$ \\
    $1.0\times 10^{-6}  $&$ 4.59\times 10^{-15} $&$ 4.05\times 10^{-9} $&$ 4.05\times 10^{-9}$ \\
    $1.0\times 10^{-7}  $&$ 4.40\times 10^{-15} $&$ 4.02\times 10^{-9} $&$ 4.02\times 10^{-9}$ \\
    $1.0\times 10^{-8}  $&$ 4.38\times 10^{-15} $&$ 4.02\times 10^{-9} $&$ 4.02\times 10^{-9}$ \\
    $1.0\times 10^{-9}  $&$ 4.38\times 10^{-15} $&$ 4.02\times 10^{-9} $&$ 4.02\times 10^{-9}$ \\
    $1.0\times 10^{-10} $&$ 4.38\times 10^{-15} $&$ 4.02\times 10^{-9} $&$ 4.02\times 10^{-9}$ \\
    \hline
    trilinear & $10^{-20...-17}$ & $10^{-20...-17}$ & $10^{-20...-17}$ \\
    bilinear  & $10^{-22...-18}$ & $10^{-22...-18}$ & $10^{-21...-18}$ \\
    \hline\hline
  \end{tabular}
\end{table}
%%%%%%%%%%%%%%%%%%%%%%%%%%%%%%%%%%%%%%%%%%%%%%%%%%%%%%%%%%%%%%%%%%%%%%%%
%
The results are shown in Tab.~\ref{tab:low} for the {\sc low} point and
Tab.~\ref{tab:high} for the {\sc high} unification point. Below the
numerical values, results for pure trilinear loops \cite{mg09} and
bilinear contributions \cite{mg18} are given as ranges, obtained for
different cases (assumptions of the normal or inverted hierarchy of
neutrino masses, data from the neutrinoless double beta decay
searches).

For the {\sc low} point the smallest contributions discussed in this
paper are at least 5 orders of magnitude greater than any other
calculated so far. This clearly shows, that the loops with heaviest
particles (like higgsinos and charginos) tend to dominate the overall
contribution to the magnetic moments. This result is not totally
unexpected, but it was not fully clear, if the higher order process
(amplitude proportional to four instead of two coupling constants) will
not suppress the effect coming directly from the masses of the heavy
particles. That was the case when the neutrino--neutralino mixing placed
two mass insertions on the external neutrino lines of the
diagrams. Here, however, the mass dependent loop functions ${\cal I}$
are capable to boost the amplitudes of the process to very high
values. All this holds also for the second, {\sc high}, set of input
parameters, although there the difference between smallest value of
$\mu_{\nu_{12}}$ and the maximum trilinear contribution is `only' two
orders of magnitude.

%%%%%%%%%%%%%%%%%%%%%%%%%%%%%%%%%%%%%%%%%%%%%%%%%%%%%%%%%%%%%%%%%%%%%%
\section{Summary}

The full contribution to the Majorana neutrino transition magnetic
moment consists of three main parts. The first one is represented by
loop diagrams with two trilinear RpV couplings, containing either
a~quark--squark or lepton--slepton pair. The second takes into account
possible neutrino--neutralino mixing, which occurs on the external
neutrino lines in the form of bilinear mass insertions. The third one,
discussed in this paper, allows for the mass insertions to appear inside
the loop, which may contain a~number of different particles, i.e.,
charged Higgs bosons and the corresponding higgsinos, leptons and
sleptons, charged gauge bosons, and charginos. A~proper discussion of
the latter contribution is possible only within a~consistent $R$-parity
violating model, in which all the mixing between different mass
eigenstates is taken into account. Also, new terms in the RGE equations,
proportional to the RpV couplings appear. However, the impact of these
terms has already been studied \cite{mg01} and their presence changed
the low energy mass spectrum of the model by a~factor of 1/5 at most
(20\%). This indicates, that the crucial point is the phenomenon of
mixing between different mass eigenstates, so that the amplitudes of the
processes must be expanded in the physical bases and summed over
respective eigenstates.

We have found that the presently discussed contributions are dominant
over the remaining two. This allowed us to find the acceptable parameter
space of the model, using the condition that no magnetic moment may
exceed $10^{-10}\mu_B$. We have also checked, how different initial
values of the RpV couplings, represented by $\Lambda_0$, change the
results. Finally, a~comparison with previous numerical studies has been
given.

%%%%%%%%%%%%%%%%%%%%%%%%%%%%%%%%%%%%%%%%%%%%%%%%%%%%%%%%%%%%%%%%%%%%%%

\section*{Acknowledgments}

This work has been financed by the Polish National Science Centre under
the decision number DEC-2011/01/B/ST2/05932.

%%%%%%%%%%%%%%%%%%%%%%%%%%%%%%%%%%%%%%%%%%%%%%%%%%%%%%%%%%%%%%%%%%%%

\end{document}